\documentclass[12pt]{iopart}

\usepackage{graphicx}
\usepackage{multirow}

\newcommand{\ve}[1]{\mbox{\boldmath$#1$}}
\newcommand{\OO}[1]{{\cal O}(c^{-#1})}
\def\source{{\rm 0}}
\def\obs{{\rm 1}} 
\newcommand{\muas}[0]{{\hbox{$\mu$as}}}
\newcommand\fr[2]{\displaystyle{\frac{#1}{#2}}}

\begin{document}
\title{On the efficient computation of the quadrupole light deflection}

\author{Sven Zschocke, Sergei A Klioner}

\address{Lohrmann Observatorium, Technische Universit\"at Dresden,
Institut f\"ur Planetare Geod\"asie, Mommsenstrasse 13, D - 01062 Dresden, Germany}
\ead{sven.zschocke@tu-dresden.de}

\begin{abstract}
  Although the formulas for the light deflection due to quadrupole
  gravitational field of deflecting bodies are well known, the
  formulas are rather complicated, so that massive computations of
  quadrupole light deflection (e.g., in the framework of astrometric
  survey missions like Gaia) are time-consuming. Considering an
  observer situated within a few million kilometers from the Earth 
  (clearly the most practical case), we derive the simplest possible form 
  of the relevant formulas still having numerical accuracy of $1\,\mu{\rm as}$. 
  This form leads to simple upper estimates for the quadrupole light deflection 
  in various cases allowing one to relate the magnitude of the actual quadrupole
  deflection with the corresponding monopole deflection due to the
  same body. These upper estimates can be used to decide if, for a
  given configuration, the actual quadrupole deflection should be
  computed for a given accuracy goal.
 \end{abstract}


\maketitle


\normalsize

\section{Introduction}

In the nearest future astrometric observations will reach an accuracy
of 1 microarcsecond (\muas) level. This level of accuracy requires a
precise modelling of light propagation. In particular, the light
deflection due to quadrupole gravitational field of deflecting bodies
should be taken into account \cite{Klioner2}.  On the other hand, the
accuracy of radio and laser radar links of future missions like
BepiColombo or Juno require modelling of the light travel time at the
level of millimeters. Therefore, also the Shapiro delay due to
quadrupole fields is of practical interest.

Analytical formulas for
quadrupole light deflection are well known.  Analytical solutions of
light deflection in a quadrupole gravitational field have been
investigated by many authors \cite{Ivanitskaya1979,Epstein_Shapiro1,
Richter_Matzner,Cowling1,Klioner1,
Klioner_Kopejkin,Klioner2,Klioner:Blankenburg:2003}. 
For the first time the full analytical
solution for the light trajectory in a quadrupole field has been
obtained in \cite{Klioner1}. These results were confirmed by a different
approach in \cite{LePoncinLafitteTeyssandier2008}. Various 
generalization (higher-order multipole moments, time-dependence, etc.) 
were derived in \cite{Kopeikin1997,KopeikinKorobkovPolnarev2006,
KopeikinMakarov2007}. The formulas suitable for high-accuracy data
reduction are given e.g. in \cite{Klioner2}. These formulas are 
rather complicated, so that massive computations of
quadrupole light deflection are time-consuming. This represents
a problem for data processing of astrometric surveys. For instance,
the ESA mission Gaia (to be launched 2012) will have to process
about $10^{12}$ individual observations of about $10^9$ distinct 
celestial objects. It is, therefore,
obvious that efficient analytical algorithms to compute the quadrupole light
deflection are mandatory. 
Furthermore, for observations in solar system, 
the quadrupole light deflection reaches the
microarcsecond level only for objects at a relatively small
angular distance from giant planets. Accordingly, it is highly useful to find
simple analytical formulas by means of which one can decide whether or not
the quadrupole field needs to be taken into account for a given accuracy and
a given geometrical configuration. 

In this paper we assume that the observer is located within a few million kilometers 
from the Earth's orbit which is clearly the most practical case (e.g. Gaia will have a
Lissajous-like orbit around Lagrange point $L_2$ of the system
Earth--Sun).  This allows one to simplify the formulas for the
quadrupole light deflection considerably. Besides these simplified formulas
we derive simple analytical estimates of the quadrupole deflection
allowing one to decide if the effect should be computed and taken into account
for a given accuracy. We also give a strict upper estimate 
for the quadrupole Shapiro delay.

The paper is organized as follows.  In Section \ref{Basics} we
summarize some basics about light deflection and introduce the
notation. In Section \ref{StarsandQuasars} the full quadrupole formula
in post-Newtonian order, the simplest possible expression for stars and
quasars still having an accuracy of $1\,\mu{\rm as}$ and 
simple upper estimates of the latter
are presented. 
In Section \ref{SolarSystemObjects} the full quadrupole formula in post-Newtonian order, 
the simplest possible expression for solar system objects still having an accuracy of $1\,\mu{\rm as}$ 
and criteria are presented. An improved estimation of quadrupole Shapiro effect is given in Section \ref{Shapiro}. 
The efficiency and correctness of the upper estimates and simplified quadrupole formulas have been investigated 
numerically and analytically, and the results are discussed in Section \ref{Numerical_Tests}. 
A summary of the findings is given in Section \ref{Summary}.


\section{Some basic formulas of light propagation}\label{Basics}

Let us summarize some basic formulas of light propagation in
post-Newtonian approximation.  The geodetic equations in
post-Newtonian order is linear with respect to the metric components
and, therefore, the coordinates of a photon and the derivative with
respect to coordinate time $t$ are given by \cite{Klioner2}
\begin{eqnarray} 
\ve{x} (t) &=& \ve{x} (t_{\source}) + 
c\,\ve{\sigma} \,(t - t_{\source}) + \sum\limits_i \Delta \ve{x}_i (t)\,+\OO4,
\label{lightpath_1}
\\
\dot{\ve{x}} (t) &=& c\,\ve{\sigma} + \sum\limits_i \Delta \dot{\ve{x}}_i (t)\,+\OO4.
\label{lightpath_2}
\end{eqnarray}
\noindent
The sum runs over individual terms in the metric of
various physical origins (e.g. monopole gravitational field of various
bodies, quadrupole fields, 
higher-order multipole fields, etc.). 
Here, $t_{\source}$ is the moment of emission,
$\ve{x}_{\source} = \ve{x} (t_{\source})$ is the position of source
and $\ve{\sigma} = \lim_{t \rightarrow - \infty} \,\fr{\dot{\ve{x}} (t)}{c}$ 
is the unit tangent vector of
light path at infinitely past. The position of observer is
$\ve{x}_{\obs} = \ve{x} (t_{\obs})$ and $t_{\obs}$ is the moment of
observation. The unit coordinate direction of the light propagation at
the moment of observation reads $\ve{n} = \fr{\dot{\ve{x}} (t_{\obs})}{\left|\dot{\ve{x}}
    (t_{\obs})\right|}$.  In post-Newtonian order the transformation
$\ve{\sigma}$ to $\ve{n}$ reads \cite{Klioner1,Klioner2}
\begin{eqnarray}
\label{lightpath_B}
\ve{n} &=& \ve{\sigma} + \sum \limits_i \delta \ve{\sigma}_i+\OO4\,,
\qquad
\delta\ve{\sigma}_i=
\ve{\sigma}\times\left(c^{-1} \Delta\dot{\ve{x}}_i\left(t_{\obs}\right)
\times\ve{\sigma}\right)\,.
\end{eqnarray}
\noindent
The spherical symmetric part (monopole contribution) due to one massive body $A$ 
and its absolute value are given by (cf. Eq.~(102) in \cite{Article_Klioner_Zschocke})
\begin{eqnarray}
\fl \delta \ve{\sigma}_{\rm pN}^A = - \left(1+\gamma\right)\,\fr{G}{c^2}\,M_A\,
\fr{\ve{d}_A}{d_A^2}\,\left(1 + \fr{\ve{\sigma}\cdot\ve{r}_1^A}{r_1^A}\right),\quad
\left|\,\delta \ve{\sigma}_{\rm pN}^A\,\right|=
\left(1+\gamma\right)\,\fr{G}{c^2}\,
\fr{M_A}{d_A}\,\left(1 + \fr{\ve{\sigma}\cdot\ve{r}_1^A}{r_1^A}\right).
\nonumber\\
\label{absolute_spherical}
\end{eqnarray}
\noindent
Here $\gamma$ is the PPN parameter, $M_A$ is the mass of body $A$, $c$ is the speed of light, 
$G$ is the gravitational constant, and 
$\ve{d}_A = \ve{\sigma} \times \left(\ve{r}_{\obs}^A \times \ve{\sigma}\right)$
is the impact parameter, $d_A = \left|\ve{d}_A\right|$, 
$\ve{r}_{\obs}^A = \ve{x} (t_{\obs}) - \ve{x}_A$, $r_1^A = \left|\ve{r}_{\obs}^A\right|$, 
and $\ve{x}_A$ is the position of massive body $A$.

In order to consider light propagation between two given points
$\ve{x}_0$ and $\ve{x}_1$ (as it is needed for the data processing for
solar system objects) let us define vectors $\ve{r}_{\source}^A =
\ve{x} (t_{\source}) - \ve{x}_A$ and $\ve{R} = \ve{x}_{\obs} -
\ve{x}_{\source} = \ve{r}_{\obs}^A - \ve{r}_{\source}^A$ with absolute
values $R = |\ve{R}|$ and $r_{\source}^A = |\ve{r}_{\source}^A|$, and
unit vector $\ve{k} = \ve{R}/R$.  In post-Newtonian approximation, the
transformation $\ve{k}$ to $\ve{n}$ reads \cite{Klioner2}
\begin{eqnarray}
\fl
\label{criterion_source_43}
\ve{n} &=& \ve{k} + \sum\limits_i \delta \ve{k}_i +\OO4 \,, 
\qquad
\delta\ve{k}_i = 
\ve{k}\times\left(\left[c^{-1} \Delta\dot{\ve{x}}_i(t_{\obs})
-R^{-1}\,\Delta\ve{x}_i(t_{\obs})\right]\times\ve{k}\right)\,.
\end{eqnarray}
\noindent
The impact parameter $\ve{d}_A$ can be computed as $\ve{d}_A = \ve{k}
\times \left(\ve{r}_{\obs}^A \times \ve{k}\right) +\OO2 = \ve{k}
\times \left(\ve{r}_{\source}^A \times \ve{k}\right)+\OO2$.
The spherical symmetric part (monopole contribution) due to one massive body $A$ 
and its absolute value are given by (cf. Eq.~(70) in \cite{Klioner2} or Eq.~(24) in \cite{Article_Klioner_Zschocke}) 
\begin{eqnarray}
\fl 
\delta \ve{k}_{\rm pN}^A = - \left(1+\gamma\right)\,\fr{G}{c^2}\,\fr{M_A}{r_{\obs}^A}\,
\fr{\ve{k} \times \left(\ve{r}_{\source}^A \times \ve{r}_{\obs}^A\,\right)}
{r_{\source}^A\,r_{\obs}^A + \ve{r}_{\source}^A \cdot \ve{r}_{\obs}^A}\,,\quad
\left|\,\delta \ve{k}_{\rm pN}^A \right|=\left(1+\gamma\right)\,\fr{G}{c^2}\, 
\fr{M_A}{r_{\obs}^A}\,\fr{\left|\,\ve{r}_{\source}^A \times \ve{r}_{\obs}^A\,\right|}
{r_{\source}^A\,r_{\obs}^A + \ve{r}_{\source}^A \cdot \ve{r}_{\obs}^A}\,.
\nonumber\\
\label{criterion_source_50}
\end{eqnarray}


\section{The quadrupole light deflection for stars and quasars}\label{StarsandQuasars}

Using the expression $\Delta\dot{\ve{x}}_{\rm Q}\left(t_{\obs}\right)$
given by Eq.~(44) of \cite{Klioner2} one gets
\cite{Klioner:Blankenburg:2003,Report0}
\begin{eqnarray}
\fl
\delta \ve{\sigma}_{\rm Q}&=&\sum_A\delta\ve{\sigma}_{\rm Q}^A\,,\qquad
\delta\ve{\sigma}_{\rm Q}^A={1+\gamma\over 2} \,\fr{G}{c^2} \, \left[ \ve{\alpha}_A^{\prime}\;
\fr{\dot{\cal U}_A}{c} + \ve{\beta}_A^{\prime}\; \fr{\dot{\cal E}_A}{c} + \ve{\gamma}_A^{\prime}\;
\fr{\dot{\cal F}_A}{c} + \ve{\delta}_A^{\prime}\;\fr{\dot{\cal V}_A}{c}\right] \,.
\label{light_35}
\end{eqnarray}
\noindent
The scalar functions and vectorial coefficients are given by 
Eqs.~(\ref{light_40}) - (\ref{light_75}) of \ref{Quadrupole1}. 
The last three terms in Eq.~(\ref{light_35}) can be estimated as
\begin{eqnarray}
T_1^A &=& {1+\gamma\over 2} \,\fr{G}{c^2}\,\left|\,
\ve{\beta}_A^{\prime}\; \fr{\dot{\cal E}_A}{c} + \ve{\gamma}_A^{\prime}\;
\fr{\dot{\cal F}_A}{c} + \ve{\delta}_A^{\prime}\;\fr{\dot{\cal V}_A}{c}\,\right|
\le 13\,\fr{G}{c^2}\,\fr{M_A\,\left|J_2^A\right|\,P_A^2}{\left(r_{\obs}^{A\;min}\right)^3}\,.
\label{light_180}
\end{eqnarray}

\noindent
Here, $P_A$ is the equatorial radius, $J_2^A$ is the second zonal harmonics of
massive body $A$, and $r_{\obs}^{A\;min}$ is the minimal distance
between massive body and observer.  The proof of this estimation is
given in \cite{Report0}.

\begin{table}
\caption{\label{table1}
Numerical parameters of the giant planets taken from \cite{Encyclopedia}. In this table the values of 
$r_{\obs}^{A\;{\rm min}}$ are given under assumption that the observer is in the vicinity of Earth's orbit. 
The value $J_2$ for the Sun is taken from \cite{J2_Sun}.}
\begin{indented}
\item[] 
\begin{tabular}{@{}lrrrrr}
\br
Parameter&Sun&Jupiter&Saturn&Uranus&Neptune\\
\mr
&&&&&\\[-10pt]
$GM_A/c^2$\  [m] & 1476. & 1.40987 & 0.42215 & 0.064473 & 0.076067\\
$J_2^A\  [10^{-3}]$    & 0.0002 & 14.697 & 16.331 & 3.516 & 3.538\\
$P_A$\  [$10^6$ m]  & 696. & 71.492 & 60.268 & 25.559 & 24.764\\
$r_{\obs}^{A\;{\rm min}}$\  [$10^{12}$ m] & 0.147 & 0.59 & 1.20 & 2.59 & 4.31\\
\mr
&&&&\\[-10pt]
$GM_A\,J_2^A\,P_A^2/c^2$\ [$10^{15}$ m$^3$] & 0.143 & 
0.106 & 0.025 & 0.000148 & 0.000165 \\
\br
\end{tabular}
\end{indented}
\end{table}

\begin{table}
\caption{\label{table2} Maximal numerical values of the neglected  
terms as given by (\ref{light_180}) and (\ref{source_176})}
\begin{indented}
\item[]
\begin{tabular}{@{}llllll}
\br
Parameter&Sun&Jupiter&Saturn&Uranus&Neptune\\
\mr
&&&&\\[-10pt]
$T_1^A$\ [$\mu{\rm as}$] & \multicolumn{5}{c}{$<10^{-6}$ for Sun and giant planets}\\
$T_2^A$\ [$\mu{\rm as}$] & $1.86 \times 10^{-3}$ & $3.26 \times 10^{-2}$ & $5.32 \times 10^{-3}$ & $8.11 \times 10^{-4}$ & $5.79 \times 10^{-5}$\\
\br
\end{tabular}
\end{indented}
\end{table}

It is well known (see Table~1 of \cite{Klioner2}) that the quadrupole
light deflection in solar system can achieve the level of $1\,\mu{\rm as}$ 
only for the giant planets (and, possibly, the Sun).  Using the
parameters in Table~\ref{table1} we obtain from (\ref{light_180})
that $T_1^A \le 10^{-6}\,\muas$ for all these bodies. Furthermore, 
numerical simulations have confirmed the correctness of the 
values given in Table~\ref{table2}. 
Accordingly, these terms in (\ref{light_180}) can
safely be neglected at the level of a microarcsecond. Accordingly, the
simplest possible expression of quadrupole light deflection for stars and quasars 
still having an accuracy of $1\,\mu{\rm as}$ and valid for 
an observer situated within a few million kilometers of the Earth orbit reads:
\begin{eqnarray}
\delta \ve{\sigma}_{\rm Q}^A &=& {1+\gamma\over 2}\,\fr{G}{c^2}\,\ve{\alpha}_A^{\prime}\,
\fr{\dot{\cal U}_A}{c} \,.
\label{light_185}
\end{eqnarray}

\noindent
The simplified formula of quadrupole light deflection
(\ref{light_185}) is still a complicated expression. In order to avoid
evaluation of this term for each object in the data reduction, a simple
criterion is needed allowing one to decide whether or not it is
necessary to compute the quadrupole light deflection for a
source. The absolute value of light deflection due to the quadrupole
field of objects $A$ can be estimated as \cite{Report0}
\begin{eqnarray}
|\,\delta \ve{\sigma}_{\rm Q}^A\,| &\le& \fr{9}{4}\,{1+\gamma\over 2}\,
\fr{G M_A}{c^2}\,
\left|J_2^A\right|\,\fr{P_A^2}{d_A^3}\,\left(1 + \fr{\ve{\sigma}\cdot\ve{r}_1^A}{r_1^A}\right).
\label{criterion_10}
\end{eqnarray}
\noindent
A comparison of (\ref{criterion_10}) with the absolute value of the
monopole deflection given by (\ref{absolute_spherical}) gives
\begin{eqnarray}
\left|\,\delta \ve{\sigma}_{\rm Q}^A\,\right| &\le& \fr{9}{8} \, \left|J_2^A\right|\,
\fr{P_A^2}{d_A^2}\,
\left|\,\delta \ve{\sigma}_{\rm pN}^A\,\right|\,.
\label{criterion_30}
\end{eqnarray}

\noindent
This estimate relates the quadrupole light deflection for stars and
quasars to the corresponding monopole deflection.  The latter is
relatively large, defined by a simple formula and usually computed for
each source and each deflecting body. In this case the estimate
(\ref{criterion_30}) can be computed at cost of two multiplications
(note that $d_A$ is known since it is used for $\delta
\ve{\sigma}_{\rm pN}^A$). In case when
$\left|\,\delta\ve{\sigma}_{\rm pN}\,\right|$ is not readily
available, one can use \cite{Report0}
\begin{eqnarray}
\left|\,\delta \ve{\sigma}_{\rm Q}^A\,\right| &\le& 
2\,(1+\gamma)\,{G M_A\over c^2}\,\left|J_2^A\right|\,\fr{P_A^2}{d_A^3} 
\label{criterion_32_A}
\\
\nonumber\\
&\le& 2\,(1+\gamma)\,{G M_A\over c^2}\,\left|J_2^A\right|\,\fr{1}{P_A}\,,
\label{criterion_32_B}
\end{eqnarray}

\noindent
where we use $d_A\ge P_A$ and (\ref{light_40}) is estimated by  
$\left|2 + 3 \cos \alpha - \cos^3 \alpha\right| \le 4$ for $\alpha$
being the angle between vectors $\ve{\sigma}$ and $\ve{r}_{\obs}^A$. 
Estimate (\ref{criterion_32_B}) coincides with Eq.~(41) of \cite{Klioner1}.


\section{The quadrupole light deflection for solar system objects\label{SolarSystemObjects}}

The quadrupole light deflection for solar system objects 
$\delta \ve{k}_{\rm Q}$ is defined by Eqs. (36)--(47) and (69) of 
\cite{Klioner2} and can be written as
\cite{Klioner:Blankenburg:2003,Report0}:
\begin{eqnarray}
\fl
\delta \ve{k}_{\rm Q} &=& 
\sum_A\delta \ve{k}_{\rm Q}^A\,,\qquad
\delta \ve{k}_{\rm Q}^A 
={1+\gamma\over 2}\,  \fr{G}{c^2}\,\sum\limits_{A} \, 
\left[\ve{\alpha}_A^{\prime \prime}
\fr{{\cal A}_A}{c} + \ve{\beta}_A^{\prime \prime}\,
\fr{{\cal B}_A}{c} + \ve{\gamma}_A^{\prime \prime}\,
\fr{{\cal C}_A}{c} + \ve{\delta}_A^{\prime \prime} \,
\fr{{\cal D}_A}{c}\right] \,.
\label{source_5}
\end{eqnarray}
\noindent
The scalar functions and vectorial coefficients are given in Eqs.~(\ref{source_10}) - (\ref{source_45}) in 
\ref{Quadrupole2}. In \cite{Report0} it has been shown that the last three terms in (\ref{source_5}) can be estimated by 
\begin{eqnarray}
T_2^A &=& {1+\gamma\over 2}\,  \fr{G}{c^2}\,
\left|\,\ve{\beta}_A^{\prime \prime}\,\fr{{\cal B}_A}{c} + \ve{\gamma}_A^{\prime \prime}\,
\fr{{\cal C}_A}{c} + \ve{\delta}_A^{\prime \prime}\,\fr{{\cal D}_A}{c}\,\right|
\nonumber\\
&\le&  \left(\fr{9}{2}\,\fr{1}{P_A^2\,r_{\obs}^{A\;{\rm min}}} 
+ \fr{1}{P_A\,\left(r_{\obs}^{A\;{\rm min}}\right)^2} + \fr{19}{2}\, 
 \fr{1}{\left(r_{\obs}^{A\;{\rm min}}\right)^3}\right)\,\fr{G M_A}{c^2}\,\left|J_2^A\right|\,P_A^2\,.
\label{source_176}
\end{eqnarray}
\noindent
Using the parameters given in Table~\ref{table1} we obtain the
numerical estimates of $T_2^A$ given in Table~\ref{table2}. Moreover,
numerical simulations have confirmed the correctness of the values for
$T_2^A$ given in Table~\ref{table2}.  In view of these numerical
values, the simplest possible form of quadrupole light deflection
(\ref{source_5}) for solar system objects with an accuracy of
$1\,\mu{\rm as}$ and valid for an observer situated within a few
million kilometers of the Earth orbit is given by
\begin{eqnarray}
\delta \ve{k}_{\rm Q}^A&=&
{1+\gamma\over 2}\, \fr{G}{c^2} \,  \,
\ve{\alpha}_A^{\prime \prime}\,\fr{{\cal A}_A}{c} \,.
\label{source_180}
\end{eqnarray}

\noindent
Again, in order to avoid unnecessary computations, one needs an efficient
way to estimate the magnitude of $\delta \ve{k}_{\rm Q}^A$. This can be
done using the following inequality \cite{Report0}
\begin{eqnarray}
|\,\delta \ve{k}_{\rm Q}^A\,| &\le&
{3\,(1+\gamma)\over 2}\,\fr{GM_A}{c^2}\,
\fr{1}{d_A^2}\, \fr{1}{r_{\obs}^A}\,\left|J_2^A\right|\,P_A^2\,
\fr{\left|\,\ve{r}_{\source}^A \times \ve{r}_{\obs}^A\,\right|}
{r_{\source}^A \,r_{\obs}^A + \ve{r}_{\source}^A \cdot \ve{r}_{\obs}^A}\,.
\label{criterion_source_40}
\end{eqnarray}

\noindent
A comparison with (\ref{criterion_source_50}) yields 
\begin{eqnarray}
\left|\,\delta \ve{k}_{\rm Q}^A\,\right| &\le& \fr{3}{2} \,\fr{P_A^2}{d_A^2}\,\left|J_2^A\right|\,
\left|\,\delta \ve{k}_{\rm pN}^A\,\right|\,.
\label{criterion_source_60}
\end{eqnarray}

\noindent
This estimate allows one to estimate the magnitude of the quadrupole
light deflection in observations of solar system objects using the
monopole deflection. As it was mentioned above, the monopole
deflection is significantly larger and should be usually calculated
for each source and each gravitating body (at least for those bodies, for which
the quadrupole deflection could be sufficiently large). If
$|\,\delta \ve{k}_{\rm pN}\,|$ has been calculated, the magnitude
of $\delta \ve{k}_{\rm Q}^A$ can be estimated at cost of three multiplications
(we note that $d_A$ is required to compute $|\,\delta \ve{k}_{\rm pN}\,|$
and can be considered as known). If $|\,\delta \ve{k}_{\rm pN}\,|$ is not
readily available, we can use \cite{Report0}
\begin{eqnarray}
\left|\,\delta \ve{k}_{\rm Q}^A\,\right| &\le& 2\,(1+\gamma)\,\fr{GM_A}{c^2}
\left|J_2^A\right|\,\fr{P_A^2}{d_A^3} 
\label{criterion_source_62_A}
\\
\nonumber\\
&\le& 2\,(1+\gamma)\,\fr{GM_A}{c^2}\,
\left|J_2^A\right|\,\fr{1}{P_A}\,,
\label{criterion_source_62_B}
\end{eqnarray}

\noindent
where in the last estimate we have used $P_A \le d_A$. 


\section{Shapiro effect for solar system objects \label{Shapiro}}

According to Eq.~(\ref{lightpath_1}), the propagation time $c\,\tau = c \left(t_{\obs} - t_{\source}\right)$ 
is given by \cite{Klioner1}:
\begin{eqnarray}
c\,\tau &=& R + c\, \sum \limits_i \delta\tau_i + {\cal O} \left(c^{-4}\right) \,,
\qquad
c\,\delta\tau_i=- \ve{k}\cdot \Delta \ve{x}_i (t_{\obs})\,.
\label{shapiro_25}
\end{eqnarray}

\noindent
The formula for the Shapiro delay due to one mass monopole with mass $M_A$ is well known:
\begin{eqnarray}
c\,\delta\tau^A_{\rm pN}=\left(1+\gamma\right)\fr{GM_A}{c^2}\,
\log \fr{r_{\source}^A + r_{\obs}^A + R}{r_{\source}^A + r_{\obs}^A - R}\,.
\label{classical_Shapiro}
\end{eqnarray}
\noindent
This monopole Shapiro delay becomes unboundedly large for growing
distance $R$ between the points of emission and observations
(although it is growing logarithmically with $R$). The quadrupole Shapiro effect 
$c\,\delta\tau_{\rm Q} = - \ve{k}\cdot \Delta\ve{x}_{\rm Q}(t_{\obs})$
is given by \cite{Klioner1}: 
\begin{eqnarray}
\fl
c \,\delta\tau_{\rm Q} = \sum\limits_A \;c \,\delta\tau_{\rm Q}^A\,,\qquad 
c\,\delta\tau_{\rm Q}^A = \fr{1 + \gamma}{2} \,\fr{G}{c^2} 
\left( \delta_A \, {\cal V}_A + \gamma_A \, {\cal F}_A + \beta_A \, {\cal E}_A \right),
\label{shapiro_50}
\end{eqnarray}

\noindent
where the scalar functions and scalar coefficients are given in
Eqs.~(\ref{shapiro_55}) - (\ref{shapiro_80}) in \ref{Quadrupole3}.   
This expression for the quadrupole Shapiro delay cannot be reasonably 
simplified. However, one can give a strict upper bound for the quadrupole effect 
in the Shapiro delay. One can demonstrate \cite{Report0} that
\begin{eqnarray}
&& \fr{G}{c^2} \left|\,\delta_A\,{\cal V}_A\,\right| \le \fr{G\,M_A}{c^2}\,\left|J_2^A\right|\,\fr{P_A^2}{d_A^2}\,,
\label{shapiro_140}
\\
&& \fr{G}{c^2} \left|\,\gamma_A\,{\cal F}_A + \beta_A\,{\cal E}_A\,\right| \le  \fr{G\,M_A}{c^2}\,\left|J_2^A\right|\,
\left(\fr{P_A^2}{\left(r_{\source}^A\right)^2} + \fr{P_A^2}{\left(r_{\obs}^A\right)^2}\right)\,.
\label{shapiro_141}
\end{eqnarray}
\noindent
Now since $d_A\ge P_A$, $r_{\source}^A\ge P_A$, and $r_{\obs}^A\ge P_A$, we conclude that
\begin{eqnarray}
\left|\,c\,\delta\tau_{\rm Q}^A\,\right| 
&\le& 3\,\left|J_2^A\right|\,\fr{GM_A}{c^2}\,,
\label{shapiro_150}
\end{eqnarray}

\noindent
which represents a strict upper bound of quadrupole Shapiro delay and slightly 
improves the estimate given in Eq.~(47) in \cite{Klioner1}.
This estimate implies that the quadrupole Shapiro delay has an upper
bound that depends only on physical parameters of the massive body.
Table~\ref{table3} gives maximal possible quadrupole effects in the
Shapiro delay for {\it any} positions of the source and observer. 

\begin{table}
\caption{\label{table3} Numerical values of estimate (\ref{shapiro_150}).}
\begin{indented}
\item[]
\begin{tabular}{@{}lrrrrr}
\br
Parameter&Sun&Jupiter&Saturn&Uranus&Neptune\\
\mr
&&&&&\\[-10pt]
$3\,\left|J_2^A\right|\,\fr{G\;M_A}{c^2}$\ [${\rm mm}$] 
& $0.89$ & $62.16$ & $20.68$ & $0.68$ & $0.81$\\
\mr
\end{tabular}
\end{indented}
\end{table}


\section{Efficiency of the upper estimates \label{Numerical_Tests}}

As explained above, the principal merit of the simple upper estimates
for the quadrupole light deflection (Eqs.~(\ref{criterion_30}),
(\ref{criterion_32_A}) and (\ref{criterion_32_B}) for stars and
quasars and Eqs.~(\ref{criterion_source_60}),
(\ref{criterion_source_62_A}) and (\ref{criterion_source_62_B}) for
solar system objects) is the possibility to use them, at very low computational cost, 
as criteria to decide if the quadrupole deflection should be calculated or not for a
given configuration and a given numerical accuracy.  In this Section
we investigate the numerical efficiency of the criteria for two
situations: (1) purely random homogeneous distribution of sources and
the position of observer with respect to the deflecting body, and (2)
light rays grazing the surface of the deflecting body, but with
directions still randomly distributed with respect to the body (note that the body is not
spherically symmetric and the orientation does play a role). For
both of these situations we compute minimal, maximal and mean values of
the ratio between the quadrupole deflection and its upper estimate. The higher
is the mean value the more efficient is the corresponding estimate as a criterion.

For stars and quasars, starting from (\ref{criterion_30}), (\ref{criterion_32_A}) and (\ref{criterion_32_B}),
we consider the following ratios 
\begin{eqnarray}
\fl r_1 = \fr{\left|\,\delta \ve{\sigma}_{\rm Q}^A\,\right|}{\fr{9}{8}\,
\fr{P_A^2}{d_A^2}\,\left|J_2^A\right|\,\left|\,\delta \ve{\sigma}_{\rm pN}^A\,\right|}\,,
\quad 
r_2 = \fr{\left|\,\delta \ve{\sigma}_{\rm Q}^A\,\right|}{4\,\fr{G\,M_A}{c^2}\,
\left|J_2^A\right|\,\fr{P_A^2}{d_A^3}}
\,,\quad
r_3 = \fr{\left|\,\delta \ve{\sigma}_{\rm Q}^A\,\right|}{4\,\fr{G\,M_A}{c^2}\,
\left|J_2^A\right|\,\fr{1}{P_A}}\,,
\label{ratio_1}
\end{eqnarray}
\noindent
where $\delta \ve{\sigma}_{\rm Q}^A$ and $\delta \ve{\sigma}_{\rm
  pN}^A$ are determined by Eqs.~(\ref{light_185}) and
(\ref{absolute_spherical}), respectively.
For solar system objects, starting from  
Eqs.~(\ref{criterion_source_60}), (\ref{criterion_source_62_A}) 
and (\ref{criterion_source_62_B}), we consider the ratios
\begin{eqnarray}
\fl r_4 = \fr{\left|\,\delta \ve{k}_{\rm Q}^A\,\right|}{\fr{3}{2}\,
\fr{P_A^2}{d_A^2}\,\left|J_2^A\right|\,\left|\,\delta \ve{k}_{\rm pN}^A\,\right|}\,,
\quad r_5 = \fr{\left|\,\delta \ve{k}_{\rm Q}^A\,\right|}{4\,\fr{G\,M_A}{c^2}\,
\left|J_2^A\right|\,\fr{P_A^2}{d_A^3}}\,,
\quad
r_6 = \fr{\left|\,\delta \ve{k}_{\rm Q}^A\,\right|}{4\,\fr{G\,M_A}{c^2}\,
\left|J_2^A\right|\,\fr{1}{P_A}}\,,
\label{ratio_2}
\end{eqnarray}
\noindent
where $\delta \ve{k}_{\rm Q}^A$ and $\delta \ve{k}_{\rm pN}^A$ is determined by Eqs.~(\ref{source_180}) and 
(\ref{criterion_source_50}), respectively. 

For all six ratios it is easy to compute the minimal and maximal
values analytically.  Besides that, for both distributions of sources
and observers it is possible to compute analytically the mathematical
expectations (i.e. the mean values) of each of six ratios $r_i$. These
values are given in Table~\ref{Numerics_1}. The analytical
calculations have been also confirmed by direct numerical simulations
in which the ratios $r_i$ were computed for correspondingly
distributed sources and positions of the observer and statistically
analyzed. 

The minimal values of all $r_i$ is zero. The maximal values of the
ratios are 1 except for $r_1$ and $r_4$ for grazing rays. In the
latter case the maximal values are less than 1. This reflects the fact
that for grazing rays the numerical coefficients in
(\ref{criterion_30}) and (\ref{criterion_source_60}) can be improved
(but only for grazing rays and not for arbitrary situation).  The fact
that no maximal values are greater than 1 confirms the validity of the
estimates. 

Note that the mean values of $r_3$ and $r_6$ for the random
distribution depend on the ratio $P_A/r_1^A$ of the equatorial radius
$P_A$ of the deflecting body and the distance between body and
observer $r_1^A$. Although this ratio can be as large as 1, it is
small in typical applications where the observer is situated far from
the body compared to the size of the latter. For Gaia the numerical
values for the mean values of $r_3$ and $r_6$ can be computed using
$P_A$ and $r_1^{A\;{\rm min}}$ from Table~\ref{table1}. For example,
for Jupiter ${1\over 3} (P_A/r_1^{A\;{\rm min}})^2=0.49\times 10^{-8}$.

The efficiency of the criteria are characterized by the
mean values of the ratios. Considering that the random distribution of
sources is much more realistic situation than the grazing rays we can
conclude that the trivial estimates (\ref{criterion_32_B}) and
(\ref{criterion_source_62_B}) leading to $r_3$ and $r_6$ are extremely
inefficient: the value of quadrupole deflection is typically many 
orders of magnitude lower than ``predicted'' by those estimates.  The
criteria (\ref{criterion_32_A}) and (\ref{criterion_source_62_A}) are
already better: the quadrupole deflection is typically only 3 times
lower than ``predicted'' (the mean value of both $r_2$ and $r_5$ for
random sources is $1/3$). It is clear, however, than the most efficient
criteria are given by (\ref{criterion_30}) and
(\ref{criterion_source_60}). For stars and quasars the value of the
quadrupole deflection ``predicted'' by (\ref{criterion_30}) is only
two times larger than the real value. It means that only 50\% of the
computations based on (\ref{criterion_30}) lead to values lower than
the desired numerical cut-off value and could be saved. Estimates
(\ref{criterion_30}) and (\ref{criterion_source_60}) will be used for
the Gaia data processing. 

\begin{table}
\caption{\label{Numerics_1}
Statistical properties of the ratios $r_i$ for two distributions of sources
(see text for further explanations).}
\begin{indented}
\item[]
\begin{tabular}{@{}cllllll}
\br
 ratio & \multicolumn{3}{c}{random} & \multicolumn{3}{c}{grazing}  \\
\mr
& min & mean  & max & \phantom{$\bigg\}$} min & mean &  max \\
$r_1$ & 0 & $\fr{40}{81}$ & 1 & \hbox to 3.5mm{} 0 & $\fr{16}{27}$ & $\fr{8}{9}$ \\
\\[-5pt]
$r_2$ & 0 & $\fr{1}{3}$ & 1 & \multirow{3}{*}{$\left.\hbox to 0cm{\vbox to 8mm{}}\right\}$\ 0} & 
\multirow{3}{*}{$\fr{2}{3}$} & \multirow{3}{*}{1} \\
\\[-5pt]
$r_3$ & 0 & $\fr{1}{3}\left({P_A/r_1^A}\right)^2$ & 1   \\
\mr
$r_4$ & 0 & $\fr{10}{27}$ & 1 & \hbox to 3.5mm{} 0 & $\fr{4}{9}$ & $\fr{2}{3}$ \\
\\[-5pt]
$r_5$ & 0 & $\fr{1}{3}$ & 1 & \multirow{3}{*}{$\left.\hbox to 0cm{\vbox to 8mm{}}\right\}$\ 0} & \multirow{3}{*}{$\fr{2}{3}$} & \multirow{3}{*}{1} \\
\\[-5pt]
$r_6$ & 0 & $\fr{1}{3}\left({P_A/ r_1^A}\right)^2$ & 1 \\
\br
\end{tabular}
\end{indented}
\end{table}


\section{Summary\label{Summary}}

In this paper we have developed efficient numerical algorithms
allowing one to compute the quadrupole light deflection
with minimal computational efforts. These algorithms will be used 
for data processing of the ESA astrometric survey mission Gaia and
can be useful in other cases. In this work we assume that the
observer is situated within a few million kilometers from the Earth orbit.
This is clearly the most practical case. Other situations can be analyzed
along the lines of our reasoning. The main results which are valid with an 
accuracy of at least $1\,\mu{\rm as}$ are as follows:
\begin{enumerate}
\item[1.] Quadrupole light deflection for stars
and quasars can be computed as (\ref{light_185}).
\item[2.] Eqs.~(\ref{criterion_30}), (\ref{criterion_32_A}) and (\ref{criterion_32_B}) 
can be used as an a priori criterion if the quadrupole light deflection 
(\ref{light_185}) has to be computed for a given source.
\item[3.] Quadrupole light deflection for solar
system sources can be computed as (\ref{source_180}). 
\item[4.] Eqs.~(\ref{criterion_source_60}), (\ref{criterion_source_62_A}) and 
(\ref{criterion_source_62_B}) can be used as an a priori criterion if the 
quadrupole light deflection (\ref{source_180}) has to be computed for a given solar system object.
\end{enumerate}
\noindent
The efficiency of the upper estimates has been investigated numerically and analytically, and the results 
are shown in Table~\ref{Numerics_1}. They demonstrate high efficiency and correctness of the upper estimates, 
both for randomly distributed sources and sources which generate grazing rays. According to these investigation, 
the most efficient upper estimate of quadrupole light deflection is (\ref{criterion_30}) for stars and quasars 
and (\ref{criterion_source_60}) for solar system objects. The correctness of simplified quadrupole formulas 
has also been shown by numerical simulations. 

Additionally, we give a strict upper bound (\ref{shapiro_150}) for 
the quadrupole effect in the Shapiro delay. This upper bound can be
used to decide whether or not the quadrupole Shapiro delay should be taken into
account if high-accuracy ranging measurements are to be modelled; 
e.g. in the framework of missions like BepiColombo \cite{BepiColumbo1} or Juno.

\ack

This work was partially supported by the BMWi grants 50\,QG\,0601 and
50\,QG\,0901 awarded by the Deutsche Zentrum f\"ur Luft- und Raumfahrt
e.V. (DLR).

\appendix

\section{Explicit formulas for the quadrupole terms}
\label{Quadrupole}

We give the full expressions of coefficients and scalar functions of quadrupole light deflection 
and quadrupole Shapiro effect in post-Newtonian order, because so far they were not being presented 
in a refereed journal.

\subsection{Light deflection for stars and quasars}
\label{Quadrupole1}

The functions in (\ref{light_35}) read
\begin{eqnarray}
\fl
\fr{\dot{\cal U}_A}{c} = \fr{1}{d_A^3} \, \left( \,2 + 3 \, \fr{\ve{\sigma} \cdot \ve{r}_{\obs}^A}{r_{\obs}^A}
- \fr{\left(\ve{\sigma} \cdot \ve{r}_{\obs}^A\right)^3}{\left(r_{\obs}^A\right)^3}\,\right) \,,
\label{light_40}
\\
\fl
\fr{\dot{\cal E}_A}{c} =
\fr{\left(r_{\obs}^A\right)^2 - 3 \, \left(\ve{\sigma}\cdot \ve{r}_{\obs}^A \right)^2}{\left(r_{\obs}^A\right)^5}\,,
\label{light_45}
\\
\fl
\fr{\dot{\cal F}_A}{c} = - 3 \; d_A \;
\fr{\ve{\sigma} \cdot \ve{r}_{\obs}^A}{\left(r_{\obs}^A\right)^5}\,,
\label{light_50}
\\
\fl
\fr{\dot{\cal V}_A}{c} = - \fr{1}{\left(r_{\obs}^A\right)^3}\,,
\label{light_55}
\end{eqnarray}
\begin{eqnarray}
\fl
\alpha_A^{'\; k} = - M_{i j}^A \, \sigma^{i} \, \sigma^j \,
\fr{d_A^k}{d_A} \, + \, 2 \,
M^A_{k j} \, \fr{d_A^{j}}{d_A}
\, - \,  2 \, M^A_{i j} \, \sigma^i \, \sigma^k \, \fr{d_A^j}{d_A}
\, - \, 4 \, M^A_{i j} \, \fr{d_A^i \, d_A^j \, d_A^k}{d_A^3} \,,
\label{light_60}
\\
\fl
\beta_{A}^{'\;k} = 2 M^A_{i j}\,\sigma^i\,\fr{d_A^j \, d_A^k}{d_A^2} \;,
\label{light_65}
\\
\fl
\gamma_A^{'\;k} = M^A_{i j} \, \fr{d_A^i \, d_A^j \, d_A^k}{d_A^3} \, - \,
M^A_{i j} \, \sigma^i \, \sigma^j \, \fr{d_A^k}{d_A} \,,
\label{light_70}
\\
\fl
\delta_A^{'\;k} = - 2 \, M^A_{i j} \, \sigma^i \, \sigma^j \, \sigma^k
\, + \, 2 \, M^A_{k j} \, \sigma^j \,
- \, 4 \, M^A_{i j} \, \sigma^i \, \fr{d_A^j \, d_A^k}{d_A^2} \,.
\label{light_75}
\end{eqnarray}

\noindent
Here $M_{ij}^A$ is the symmetric and trace-free quadrupole moment of
body $A$. For an axial symmetric body (this approximation is
sufficient for the solar system and the accuracy of $1\,\mu{\rm as}$) one has
\begin{eqnarray}
\fl
M^A_{i j} = \fr{1}{3}\;M_A\;P_A^2\;J_2^A\;{\cal R}_A\;{\rm diag}\left(1,1,-2\right)\;{\cal R}_A^{T}\,,
\label{light_95}
\end{eqnarray}

\noindent
where ${\cal R}_A$ is the rotational matrix giving the orientation of the figure 
axis of body $A$ (see Eq. (48)--(53) of \cite{Klioner2}).

\subsection{Light deflection for solar system objects}
\label{Quadrupole2}

The functions in (\ref{source_5}) read
\begin{eqnarray}
\fl
\fr{{\cal A}_A}{c} =  \fr{1}{d_A} \, \fr{1}{R}
\left(\fr{1}{r_{\source}^A} \fr{r_{\source}^A + \ve{k} \cdot \ve{r}_{\source}^A}
{r_{\source}^A - \ve{k} \cdot \ve{r}_{\source}^A}
- \fr{1}{r_{\obs}^A} \fr{r_{\obs}^A + \ve{k} \cdot \ve{r}_{\obs}^A}{r_{\obs}^A - \ve{k} \cdot \ve{r}_{\obs}^A} \right)
+ \fr{1}{d_A^3} \left(2 + 3\,\fr{\ve{k} \cdot \ve{r}_{\obs}^A}{r_{\obs}^A}
- \fr{\left(\ve{k}\cdot\ve{r}_{\obs}^A\right)^3}{\left(r_{\obs}^A\right)^3}\right),
\nonumber\\
\label{source_10}
\\
\fl 
\fr{{\cal B}_A}{c} = \fr{1}{R}
\left(\fr{\ve{k} \cdot \ve{r}_{\source}^A} {\left(r_{\source}^A\right)^3}
- \fr{\ve{k} \cdot \ve{r}_{\obs}^A} {\left(r_{\obs}^A\right)^3} \right)
+ \fr{\left(r_{\obs}^A\right)^2 - 3 \, \left(\ve{k} \cdot \ve{r}_{\obs}^A\right)^2}{\left(r_{\obs}^A\right)^5}\;,
\label{source_15}
\\
\fl \fr{{\cal C}_A}{c} = \fr{d_A}{R}
\left(\fr{1}{\left(r_{\source}^A\right)^3} - \fr{1}{\left(r_{\obs}^A\right)^3} \right) - 3 \, d_A \,
\fr{\ve{k} \cdot \ve{r}_{\obs}^A}{\left(r_{\obs}^A\right)^5}\,,
\label{source_20}
\\
\fl
\fr{{\cal D}_A}{c} =  - \fr{1}{d_A^2} \; \fr{1}{R}
\left(\fr{\ve{k}\cdot\ve{r}_{\source}^A} {r_{\source}^A} - \fr{\ve{k}\cdot\ve{r}_{\obs}^A} {r_{\obs}^A} \right)
- \fr{1}{\left(r_{\obs}^A\right)^3}\,,
\label{source_25}
\end{eqnarray}
\begin{eqnarray}
\fl 
\alpha_A^{\prime\prime\; k} = - M_{i j}^A \, k^{i} \, k^j \, \fr{d_A^k}{d_A} \,
\, + \, 2 \,
M^A_{k j} \, \fr{d_A^j}{d_A}
\, - \,  2 \, M^A_{i j} \, k^i \, k^k \, \fr{d_A^j}{d_A}
\, - \, 4 \, M^A_{i j} \, \fr{d_A^i \, d_A^j \, d_A^k}{d_A^3} \,,
\label{source_30}
\\
\fl
\beta_{A}^{\prime\prime\;k} = 2 M^A_{i j} \, k^i \,
\fr{d_A^j \, d_A^k}{d_A^2} \,,
\label{source_35}
\\
\fl
\gamma_A^{\prime\prime\;k} = M^A_{i j} \, \fr{d_A^i \, d_A^j \, d_A^k}{d_A^3}
\, - \, M^A_{i j} \, k^i \, k^j \, \fr{d_A^k}{d_A} \,,
\label{source_40}
\\
\fl 
\delta_A^{\prime\prime\;k} = - 2 \, M^A_{i j} \, k^i \, k^j \, k^k
\, + \, 2 \, M^A_{k j} \, k^j \,
- \, 4\, M^A_{i j} \, k^i \, \fr{d_A^j \, d_A^k}{d_A^2} \,.
\label{source_45}
\end{eqnarray}

\subsection{Shapiro delay}
\label{Quadrupole3}

The functions in (\ref{shapiro_50}) read
\begin{eqnarray}
\fl 
{\cal E}_A = \fr{\ve{k} \cdot \ve{r}_{\source}^A}{\left(r_{\source}^A\right)^3}
- \fr{\ve{k} \cdot \ve{r}_{\obs}^A}{\left(r_{\obs}^A\right)^3}\,,
\label{shapiro_55}
\\
\fl
{\cal F}_A = d_A \left( \fr{1}{\left(r_{\source}^A\right)^3} - \fr{1}{\left(r_{\obs}^A\right)^3} \right) \,,
\label{shapiro_60}
\\
\fl {\cal V}_A = - \fr{1}{d_A^2}
\left( \fr{\ve{k} \cdot \ve{r}_{\source}^A}{r_{\source}^A} - \fr{\ve{k} \cdot \ve{r}_{\obs}^A}{r_{\obs}^A} \right) \,,
\label{shapiro_65}
\end{eqnarray}
\begin{eqnarray}
\fl 
\beta_A = M_{i j}^A\,k_i\,k_j - M_{i j}^A\,\fr{d_A^i}{d_A}\,\fr{d_A^j}{d_A} \,,
\label{shapiro_70}
\\
\fl 
\gamma_A = 2\,M_{i j}^A\,k^i\,\fr{d_A^j}{d_A} \,,
\label{shapiro_75}
\\
\fl 
\delta_A = M_{i j}^A \,k_i\,k_j + 2\,M_{i j}^A\,\fr{d_A^i}{d_A}\,\fr{d_A^j}{d_A}\,.
\label{shapiro_80}
\end{eqnarray}

\end{document}